\documentclass[11pt,oneside,letterpaper]{article}
\usepackage{amssymb}
\usepackage{amsmath}
\usepackage[dvips]{graphicx}
\usepackage{setspace}
\usepackage{graphicx}

\begin{document}

\vspace*{2.5cm}
\begin{center}
{ \Large Exact Tunneling Solutions in Minkowski Spacetime \\ and a Candidate for Dark Energy\\}
 \vspace*{1.7cm}
Georgios Pastras$^1$\\
%\vspace*{1.7cm}
\end{center}
\vspace*{0.6cm}
\begin{center}
$^1$SB ITP LPPC, \'Ecole Polytechnique F\'ed\'erale de Lausanne, CH-1015, Lausanne, Switzerland

\vspace*{0.8cm}
\end{center}
\vspace*{1.5cm}
\doublespacing
\begin{abstract}
We study exact tunneling solutions in scalar field theory for potential barriers composed of linear or quadratic patches. We analytically continue our solutions to imaginary Euclidean radius in order to study the profile of the scalar field inside the growing bubble. We find that generally there is a non-trivial profile of the scalar field, generating a stress-energy tensor, that depending on the form of the potential, can be a candidate for dark energy.
\end{abstract}

\newpage
\setcounter{page}{1}
\pagenumbering{arabic}

\baselineskip=18pt
\numberwithin{equation}{section}

\tableofcontents
\setcounter{tocdepth}{2}

%\cfoot{\thepage}
\onehalfspacing
\newpage

\section{Introduction}

In quantum field theory containing scalars it may occur that there are more than one local minima in configuration space. In such cases a system trapped in a metastable vacuum will decay towards a vacuum with lower energy through quantum tunneling. Semiclassical methods can be used to describe the procedure \cite{Coleman:1977py}, \cite{Callan:1977pt}. In this approach the theory is studied in Euclidean space and a classical tunneling solution that matches the appropriate boundary conditions is constructed. Typically this solution describes a bubble, which separates the true vacuum from the false vacuum, and which after its emission starts expanding, asymptotically with the speed of light.

Typically it is very difficult to find analytic solutions for any given potential that contains a metastable vacuum. The original papers \cite{Coleman:1977py}, \cite{Callan:1977pt} focus on a limit, in which the energy difference between the true and false vacua is small. In this limit the radius of the emitted bubble is large in comparison to its width, thus the name ``Thin Wall Approximation" for this approach. This limit allows for analytical expressions of the bubble emission rates. However such a limit destroys all other potentially interesting features of the solution, especially in the interior of the bubble.

In order to describe potential barriers that are not appropriate for the thin wall approximation, we can either make a numerical computation, or approximate the potential with another one that is exactly solvable. The latter is analyzed in \cite{Duncan:1992ai}, where a linear and a rectangular potential are analytically solved and the relevant decay rates calculated. However, if the actual potential is smooth, such potentials are clearly not a very good approximation for the regions of the two vacua and the top of the barrier, so several qualitative features of the solution may be lost. For example, the discontinuity of the potential in the rectangular approximation removes all dynamics as the field rolls towards the true vacuum.

In this paper, we will extend the study of the triangular model and moreover solve some more realistic, still analytically solvable potentials, and try to extract qualitative features of the related physics, and possible cosmological implications. It is going to turn out that the triangular model with parameters of Planck scale can provide an elegant explanation for the order of magnitude of the measured dark energy density in our universe. Several other options with more singular potential also exist predicting the correct order of magnitude.

\section{Framework}

We are going to study exact tunneling solutions in the simple case of a single scalar field and a potential containing a unique false vacuum. We are particularly interested in analytically continuing our solutions to imaginary Euclidean radius in order to discover the evolution of the field inside the bubble. The Lagrangian describing our system is
\begin{equation}
\mathcal{L} = \frac{1}{2}\left( {\partial _\mu  \phi } \right)^2  - V\left( \phi  \right).
\end{equation}

From now on, we name the positions of the true and false vacua as $\phi_-$ and $\phi_+$ respectively and the relevant values of the potential $V_-$ and $V_+$.

It has been proven that in scalar field theory the spherically symmetric solutions are favored \cite{Coleman:1977kc}. This is intuitively reasonable, as the sphere maximizes the ratio of volume to surface. Thus we assume that the tunneling solution depends only on the Euclidean radius $\rho$.

Under the assumption that the solution depends only on the Euclidean radius, the Euclidean field equation is reduced to
\begin{equation}
\ddot \phi  + \frac{3}{\rho}\dot \phi  = V'\left( \phi  \right),
\end{equation}
where the dot implies differentiation with respect to $\rho$ and the prime implies differentiation with respect to the field $\phi$. The solution has to obey the following boundary conditions
\begin{equation}
\mathop {\lim }\limits_{\rho \to \infty } \phi \left( \rho \right) = \phi _ +  ,\quad \dot \phi \left( 0 \right) = 0.
\end{equation}
In the following we are going to make simple assumptions for the form of the potential, that are going to allow us to find analytic solutions.

\subsection{Two Kinds of Solutions}
\label{subsec:kinds}

Before we proceed to solve the equation of motion, we would like to make a comment on the general form of the solutions. Typically we are going to assume that the potential is described by different formulas before and after the top of the barrier. One would expect that we would result in a solution of the form
\begin{equation}
 \phi  =
 \begin{cases}
 \phi_- , & \rho  < R_- \\
 \phi_1 \left( \rho \right) , & R_- < \rho < R_T \\
 \phi_2 \left( \rho \right) , & R_T < \rho < R_+ \\
 \phi_+ , & \rho > R_+ \\
 \end{cases}.
\end{equation}
This describes a bubble whom profile is as following. Outside a certain radius $R_+$ the field rests in the false vacuum. Inside this radius the field climbs the barrier between $R_+$ and $R_T$ and then rolls down to the true vacuum between $R_T$ and $R_-$ and then stays there. In such cases the analytic continuation to imaginary Euclidean radius is trivially $\phi=\phi_-$. In \cite{Duncan:1992ai} we see that in the rectangular approximation, tunneling solutions always look like that, however in the triangular approximation we may get such a solution or not depending on the parameters of the potential.

However in the triangular approximation, and as we will show later on in other cases, it may be true that the field never reaches the true vacuum in Euclidean space. The conditions leading to such a result in the triangular approximation are not very restricting on the parameters of the potential. In such cases the solution is going to be of the form
\begin{equation}
 \phi  =
 \begin{cases}
 \phi_1 \left( \rho \right) , & \rho < R_T \\
 \phi_2 \left( \rho \right) , & R_T < \rho < R_+ \\
 \phi_+ , & \rho > R_+ \\
 \end{cases}
\end{equation}
and the analytic continuation to imaginary proper time is nontrivial. We will see that actually the field in such cases never reaches the true vacuum, but performs a damped oscillation around it. The rest of the paper focuses in this category of solutions.

\section{A Volcanic Potential and the Field in the Interior of the Bubble}

\subsection{The Approximation}

The only selections of potential that preserve the linearity of the equation of motion is a linear and a quadratic one. The quadratic naturally is the most obvious selection to approximate the region of a vacuum, thus we will start our analysis studying a potential barrier built out of quadratics. The simplest possible barrier potential built by quadratics is described by
\begin{equation}
 V\left( \phi  \right) =
 \begin{cases}
 \cfrac{1}{2}m_ +  ^2 \left( {\phi  - \phi _ +  } \right)^2  + V_ + , & \phi < \phi_T \\
 \cfrac{1}{2}m_ -  ^2 \left( {\phi  - \phi _ -  } \right)^2  + V_ - , & \phi > \phi_T \\
 \end{cases}
 \label{eq:volcpotential}
\end{equation}

\begin{figure}[h]
\begin{center}
\includegraphics[angle=0,width=0.65\textwidth]{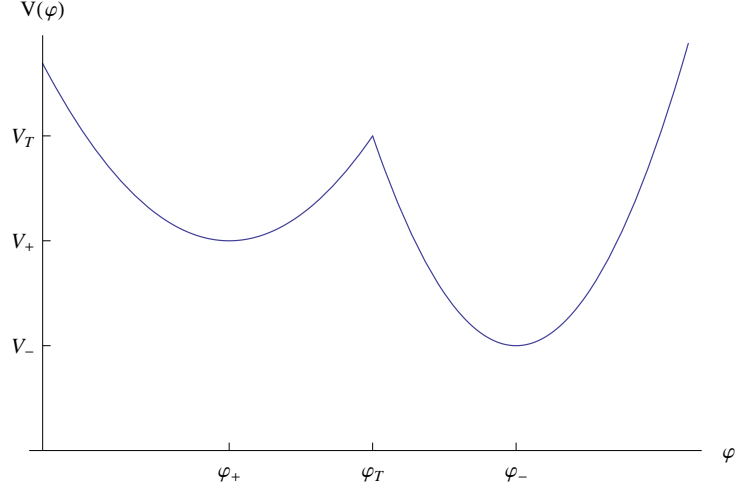}
\end{center}
\caption{The volcanic approximation}
\label{fig:quadpot}
\end{figure}

and looks like in figure \ref{fig:quadpot}. Because of the shape of such a potential, we call this the volcanic approximation.

\subsection{The Instanton Solution}

In order to find the tunneling solution we need to find the general solution to the equation
\begin{equation}
\ddot \phi  + \frac{3}{\rho }\dot \phi  = m^2 \left( {\phi  - \phi _0 } \right).
\end{equation}
If we make the substitution
\begin{equation}
\phi  - \phi _0  = \frac{y}{\rho },
\end{equation}
the equation is written as
\begin{equation}
\rho ^2 \ddot y + \rho \dot y - \left( {m^2 \rho ^2  + 1} \right)y = 0,
\end{equation}
which is exactly the modified Bessel equation, $x^2 y'' + x y' - \left( {x^2  + \alpha^2} \right)y = 0$ for $x=m\rho$ and $\alpha=1$. Thus the general solution is
\begin{equation}
\phi  = \phi _0  + \frac{{c_1 I_1 \left( {m\rho } \right) + c_2 K_1 \left( {m\rho } \right)}}{\rho },
\end{equation}
where $I$ is the modified Bessel function of the first kind and $K$ the modified Bessel function of the second kind. In the construction of the instanton solution we are going to need the derivative of the solution. This is given by
\begin{equation}
\dot \phi  = m \frac{{c_1 I_2 \left( {m\rho } \right) - c_2 K_2 \left( {m\rho } \right)}}{\rho }.
\end{equation}

Using the above result it is clear that a solution that does not reach the true vacuum in Euclidean space will look like
\begin{equation}
 \phi  =
 \begin{cases}
 \phi _ -   + \cfrac{{c_{1 - } I_1 \left( {m_ -  \rho } \right) + c_{2 - } K_1 \left( {m_ -  \rho } \right)}}{\rho } , & \rho < R_T \\
 \phi _ +   + \cfrac{{c_{1 + } I_1 \left( {m_ +  \rho } \right) + c_{2 + } K_1 \left( {m_ +  \rho } \right)}}{\rho } , & R_T < \rho < R_+ \\
 \phi_+ , \phantom{\cfrac{{1}}{{2}}} & \rho > R_+ \\
 \end{cases}.
\end{equation}
Let's now apply the boundary and matching conditions to specify the undetermined constants and the radii. The solution has to be stationary at the origin. As $K_1$ diverges at the origin, and $I_1$ is stationary, we get
\begin{equation}
c_{2 - }  = 0.
\end{equation}
Demanding continuity of the solution and its derivative at $\rho=R_+$ gives us the following two equations
\begin{align}
c_{1 + } I_1 \left( {m_ +  R_ +  } \right) + c_{2 + } K_1 \left( {m_ +  R_ +  } \right) &= 0, \\
c_{1 + } I_2 \left( {m_ +  R_ +  } \right) - c_{2 + } K_2 \left( {m_ +  R_ +  } \right) &= 0.
\end{align}
As both modified Bessel functions of the first and second kind are positive, the only solution to this problem for any finite $R_+$ is $c_{1 + }=c_{2 + }=0$. However as modified Bessel functions of the second kind decrease exponentially at infinity, we have the option that actually $R_+$ is infinite and
\begin{equation}
c_{1 + }  = 0.
\end{equation}
Demanding that $\mathop {\lim }\limits_{\rho \to {R_T}^- } \phi \left( \rho \right) = \mathop {\lim }\limits_{\rho \to {R_T}^+ } \phi \left( \rho \right) = \phi _ T$ gives us
\begin{align}
c_{1 - }  &=  - \frac{{\phi _ -   - \phi _T }}{{I_1 \left( {m_ -  R_T } \right)}}R_T,\\
c_{2 + }  &=  \frac{{\phi _ T   - \phi _+ }}{{K_1 \left( {m_ +  R_T } \right)}}R_T,
\end{align}
We have expressed all parameters in terms of $R_T$. Finally demanding continuity of the derivative at $\rho=R_T$, specifies this.
\begin{equation}
\frac{{K_1 \left( {m_ +  R_T } \right)}}{{K_2 \left( {m_ +  R_T } \right)}} \frac{{I_2 \left( {m_ -  R_T } \right)}}{{I_1 \left( {m_ -  R_T } \right)}} = \frac{{m_ +}}{{m_ -}} \frac{{\phi _ T   - \phi _+ }}{{\phi _ -   - \phi _T }}.
\label{eq:volcnum}
\end{equation}
This equation is not analytically solvable, so we cannot acquire an analytic expression for $R_T$.

To sum up the tunneling solution is
\begin{equation}
 \phi  =
 \begin{cases}
 \phi _ -   - \cfrac{{R_T \left( {\phi _ -   - \phi _T } \right)}}{\rho }\cfrac{{I_1 \left( {m_ -  \rho } \right)}}{{I_1 \left( {m_ -  R_T } \right)}} , & \rho < R_T \\
 \phi _ +   + \cfrac{{R_T \left( {\phi _ T   - \phi _+ } \right)}}{\rho }\cfrac{{K_1 \left( {m_ +  \rho } \right)}}{{K_1 \left( {m_ +  R_T } \right)}} , & \rho > R_T \\
 \end{cases},
\end{equation}
where $R_T$ is given by (\ref{eq:volcnum}).

\subsection{Condition for a Dumping Instanton}

\begin{figure}[h]
\begin{center}
\includegraphics[angle=0,width=0.45\textwidth]{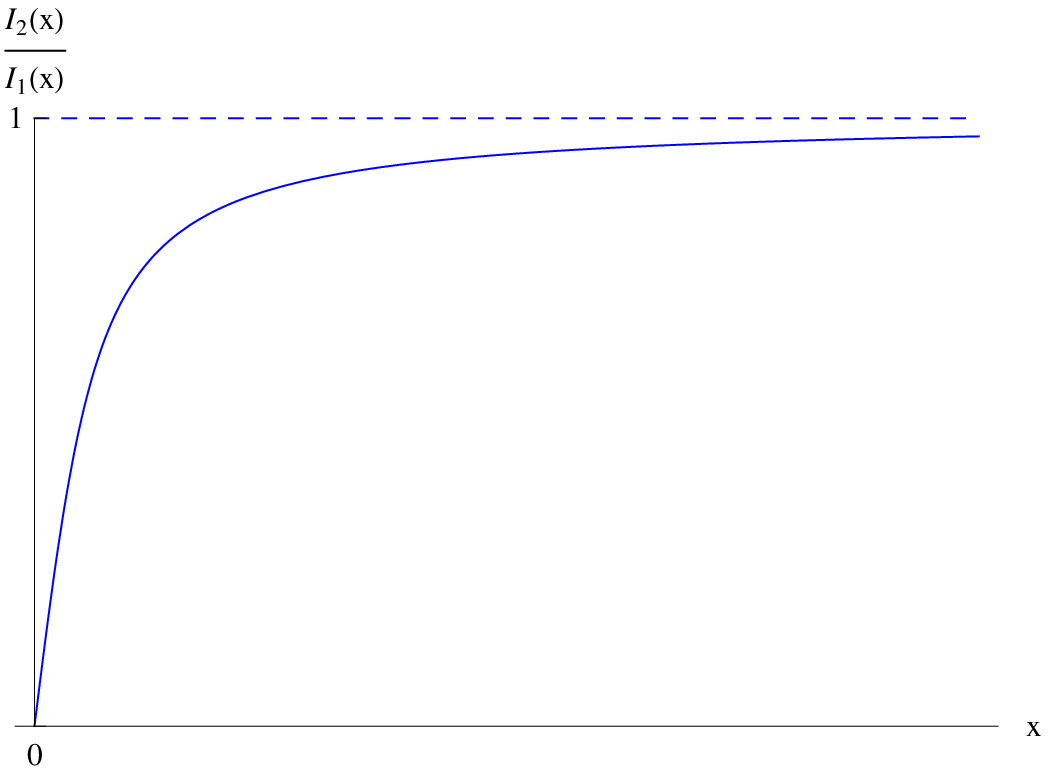} \includegraphics[angle=0,width=0.45\textwidth]{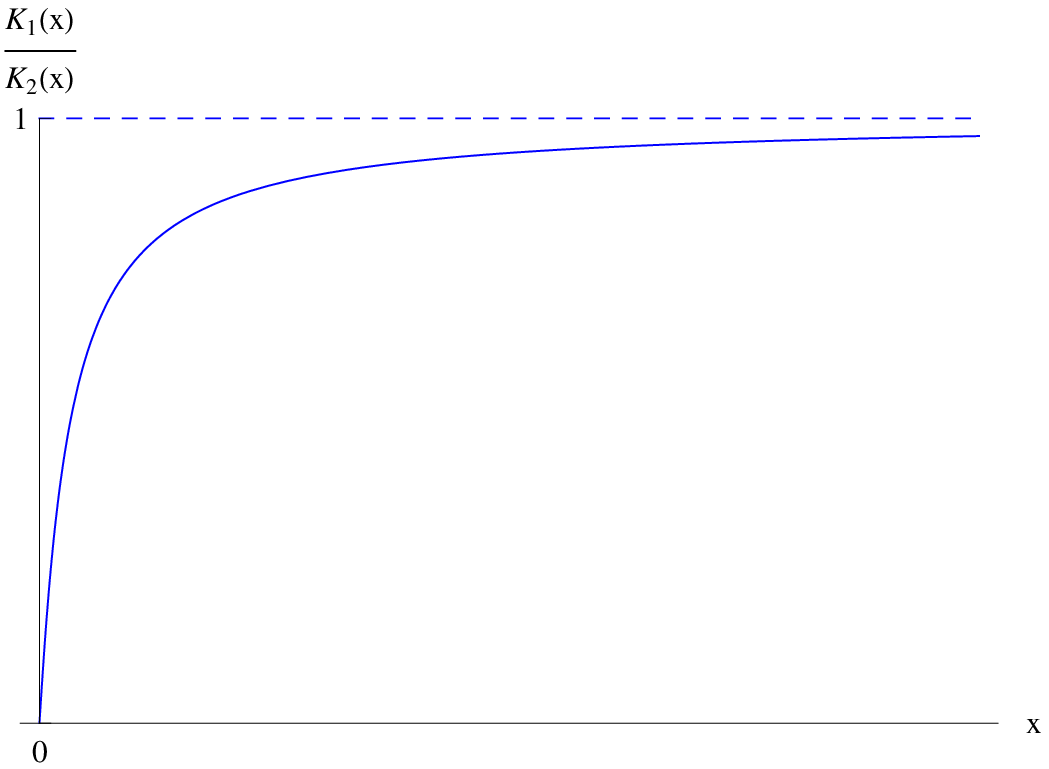}
\end{center}
\caption{The numerical part of the solution}
\label{fig:volcanocondition}
\end{figure}

We expect that in analogy to the triangular approximation \cite{Duncan:1992ai}, if equation (\ref{eq:volcnum}) does not have a solution, the solution reaches the true vacuum in Euclidean space. Both functions $\frac{{K_1 \left( {x} \right)}}{{K_2 \left( {x} \right)}}$ and $\frac{{I_2 \left( {x} \right)}}{{I_1 \left( {x} \right)}}$ are monotonous positive and take values between zero and one, as one can see in figure \ref{fig:volcanocondition}.

Thus there is always exactly one solution, as long as
\begin{equation}
\frac{{m_ +}}{{m_ -}} \frac{{\phi _ T   - \phi _+ }}{{\phi _ -   - \phi _T }} < 1,
\label{eq:volccondition}
\end{equation}
otherwise we should expect a solution that reaches the true vacuum in Euclidean space. However, such a solution will look like
\begin{equation}
 \phi  =
 \begin{cases}
 \phi_- , \phantom{\cfrac{{1}}{{2}}} & \rho < R_- \\
 \phi _ -   + \cfrac{{c_{1 - } I_1 \left( {m_ -  \rho } \right) + c_{2 - } K_1 \left( {m_ -  \rho } \right)}}{\rho } , & R_- < \rho < R_T \\
 \phi _ +   + \cfrac{{c_{1 + } I_1 \left( {m_ +  \rho } \right) + c_{2 + } K_1 \left( {m_ +  \rho } \right)}}{\rho } , & R_T < \rho < R_+ \\
 \phi_+ , \phantom{\cfrac{{1}}{{2}}} & \rho > R_+ \\
 \end{cases}.
\end{equation}
Demanding continuity and smoothness at $\rho=R_-$, gives us
\begin{align}
c_{1 - } I_1 \left( {m_ -  R_ -  } \right) + c_{2 - } K_1 \left( {m_ -  R_ -  } \right) &= 0, \\
c_{1 - } I_2 \left( {m_ -  R_ -  } \right) - c_{2 - } K_2 \left( {m_ -  R_ -  } \right) &= 0.
\end{align}
As the modified Bessel function are positive, the only solution to the above system of equations for any $R_->0$ is $c_{1 - }=c_{2 - }=0$. The only way to save this is to set $R_-$ to zero. Then smoothness at $\rho=R_-$ gives us $c_{2 - }=0$. However then the field has not reached the true vacuum at $\rho=R_-$, as $\mathop {\lim }\limits_{x \to 0} \frac{{I_1 \left( x \right)}}{x} = \frac{1}{2} \ne 0$.

Thus it looks like finding a solution that reaches the true vacuum in Euclidean space is problematic. Indeed we can see that condition (\ref{eq:volccondition}) is always satisfied. From the expression of the potential (\ref{eq:volcpotential}) we can find
\begin{equation}
\phi _T  = \frac{{m_ +  \phi _ +   - m_ -  \phi _ -   + \sqrt {m_ +  m_ -  \left( {\phi _ -   - \phi _ +  } \right)^2  + 2\left( {m_ +   - m_ -  } \right)\left( {V_ +   - V_ -  } \right)} }}{{m_ +   - m_ -  }}.
\end{equation}
Thus
\begin{multline}
\frac{{m_ +  }}{{m_ -  }}\frac{{\phi _T  - \phi _ -  }}{{\phi _ +   - \phi _T }} \\
= \frac{{m_ +  }}{{m_ -  }}\frac{{m_ -  ^2 \left( {\phi _ -   - \phi _ +  } \right) - \sqrt {m_ +  ^2 m_ -  ^2 \left( {\phi _ -   - \phi _ +  } \right)^2  - 2\left( {m_ +  ^2  - m_ -  ^2 } \right)\left( {V_ +   - V_ -  } \right)} }}{{m_ +  ^2 \left( {\phi _ -   - \phi _ +  } \right) + \sqrt {m_ +  ^2 m_ -  ^2 \left( {\phi _ -   - \phi _ +  } \right)^2  - 2\left( {m_ +  ^2  - m_ -  ^2 } \right)\left( {V_ +   - V_ -  } \right)} }},
\end{multline}
which implies
\begin{multline}
\left( {\frac{{m_ +  }}{{m_ -  }}\frac{{\phi _T  - \phi _ -  }}{{\phi _ +   - \phi _T }}} \right)^2  \\
= \frac{{1 - 2\frac{{m_ +  ^4 \left( {V_ +   - V_ -  } \right) + m_ +  ^2 m_ -  ^2 \left( {\phi _ -   - \phi _ +  } \right)\sqrt {m_ +  ^2 m_ -  ^2 \left( {\phi _ -   - \phi _ +  } \right)^2  - 2\left( {m_ +  ^2  - m_ -  ^2 } \right)\left( {V_ +   - V_ -  } \right)} }}{{m_ +  ^2 m_ -  ^2 \left( {m_ +  ^2  + m_ -  ^2 } \right)\left( {\phi _ -   - \phi _ +  } \right)^2  + 2m_ +  ^2 m_ -  ^2 \left( {V_ +   - V_ -  } \right)}}}}{{1 + 2\frac{{m_ -  ^4 \left( {V_ +   - V_ -  } \right) + m_ +  ^2 m_ -  ^2 \left( {\phi _ -   - \phi _ +  } \right)\sqrt {m_ +  ^2 m_ -  ^2 \left( {\phi _ -   - \phi _ +  } \right)^2  - 2\left( {m_ +  ^2  - m_ -  ^2 } \right)\left( {V_ +   - V_ -  } \right)} }}{{m_ +  ^2 m_ -  ^2 \left( {m_ +  ^2  + m_ -  ^2 } \right)\left( {\phi _ -   - \phi _ +  } \right)^2  + 2m_ +  ^2 m_ -  ^2 \left( {V_ +   - V_ -  } \right)}}}} < 1.
\end{multline}
That means that condition (\ref{eq:volccondition}) always holds. Thus in the volcanic approximation, there is always exactly one tunneling solution, that never reaches the true vacuum in Euclidean space. This implies that we should expect to find some kind of damped oscillation around the true vacuum in the interior of the bubble, if we analytically expand to imaginary Euclidean radius for any parameters of the volcanic potential. As this behavior is determined by the potential at the region of the true vacuum, the fact that the solution in the case of the triangular or rectangular approximation may be constant for imaginary Euclidean distance \cite{Duncan:1992ai}, is an effect because of the non-smoothness of the potential. We expect that any smooth potential produces tunneling solution with the characteristic behavior of the damped oscillation around the true vacuum in the interior of the bubble.

\subsection{The Analytical Continuation to Lorentzian Spacetime}

\begin{figure}[h]
\begin{center}
\includegraphics[angle=0,width=0.65\textwidth]{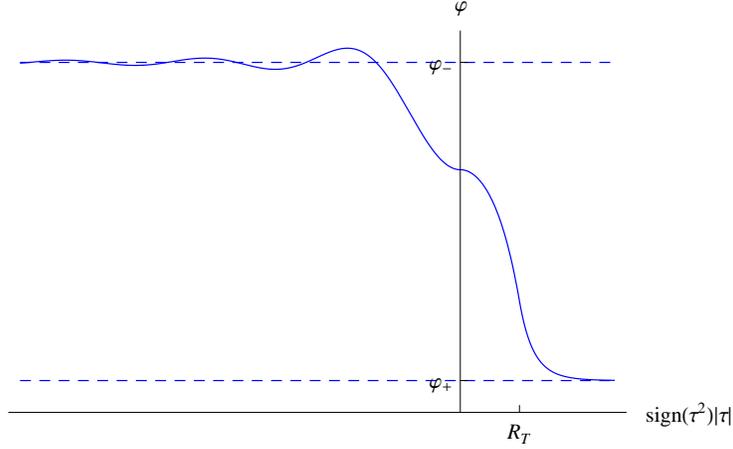}
\end{center}
\caption{The Instanton Solution}
\label{fig:volcsol}
\end{figure}

The solution of the volcanic potential is very easy to analytically continue to imaginary proper time. The modified Bessel functions are analytic functions having the property $I_1 \left( x \right) =  - iJ_1 \left( {ix} \right)$, where $J$ is the Bessel function of the first kind. Thus for $\rho=i \tau$ we get
\begin{equation}
\phi \left( \tau  \right) = \phi _ -   - \frac{{R_T \left( {\phi _ -   - \phi _T } \right)}}{\tau }\frac{{J_1 \left( {m_ -  \tau } \right)}}{{I_1 \left( {m_ -  R_T } \right)}},
\end{equation}
which clearly describes a damped oscillation of the field around the true vacuum in the interior of the bubble. For large $\tau$ we can use the asymptotic formula for the Bessel function to get
\begin{equation}
\phi \left( \tau  \right) \simeq \phi _ -   + \sqrt {\frac{2}{\pi }} \frac{{R_T \left( {\phi _ -   - \phi _T } \right)}}{{I_1 \left( {m_ -  R_T } \right)}}\frac{{\cos \left( {m_ -  \tau  + \frac{\pi }{4}} \right)}}{{\tau ^{\frac{3}{2}} }}.
\end{equation}
The solution is plotted in figure \ref{fig:volcsol}.

\subsection{The Stress-Energy Tensor in the Interior of the Bubble}

For possible applications to cosmology, it would be interesting to calculate the stress-energy tensor in the interior of the bubble. Unlike traditional treatment, where the stress energy tensor contains only the vacuum energy of the true vacuum, here the damped oscillation around it is going to contribute too.

The stress-energy tensor is given by
\begin{equation}
{T^\mu}  _\nu   = \frac{{\partial \mathcal{L}}}{{\partial \left( {\partial _\mu  \phi } \right)}}\partial _\nu  \phi  - \mathcal{L} \delta ^\mu  _\nu.
\end{equation}
Applying our Lagrangian we get
\begin{equation}
{T^\mu}  _\nu   = \partial ^\mu  \phi \partial _\nu  \phi  - \mathcal{L} \delta ^\mu  _\nu.
\end{equation}

Our solution depends only on proper time, thus
\begin{equation}
\partial _\nu  \phi  = \dot \phi \left( \tau  \right)\frac{{x_\nu  }}{{\sqrt { - x_\mu  x^\mu  } }}.
\end{equation}
Using the above we get
\begin{equation}
{T^\mu}  _\nu   = \dot \phi ^2 \frac{{x^\mu  x_\nu  }}{{ x_\lambda  x^\lambda  }} -\mathcal{L} \delta ^\mu  _\nu.
\end{equation}
The first term of the stress-energy tensor is the kinetic energy of the remnant of the bubble wall inside the bubble. Far away from the bubble wall, it contributes only to the time-time component of the stress-energy tensor, thus it behaves as dark matter. The second term is identical to vacuum energy, that does not only originate from the energy of the true vacuum, but also on the oscillation of the field.
\begin{align}
\rho _{vac}  &= \frac{1}{2}\left( {- \dot \phi ^2  + m_ -  ^2 \left( {\phi  - \phi _ -  } \right)^2 } \right) + V _ - ,\\
\rho _{wall}  &= \dot \phi ^2.
\end{align}
Using our solution we get
\begin{align}
\rho _{vac}  &= \frac{{m_ -  ^2 R_T ^2 \left( {\phi _ -   - \phi _T } \right)^2 }}{{2I_1 \left( {m_ -  R_T } \right)^2 \tau ^2 }}\left( {- J_1 \left( {m_ -  \tau } \right)^2  + J_2 \left( {m_ -  \tau } \right)^2 } \right) + V _ - , \\
\rho _{wall}  &= \frac{{m_ -  ^2 R_T ^2 \left( {\phi _ -   - \phi _T } \right)^2 }}{{I_1 \left( {m_ -  R_T } \right)^2 \tau ^2 }}J_2 \left( {m_ -  \tau } \right)^2.
\end{align}
For large $\tau$ we can use the asymptotic formula for the Bessel function to get
\begin{align}
\rho _{vac}  &= \frac{{m_ -  R_T ^2 \left( {\phi _ -   - \phi _T } \right)^2 }}{{\pi I_1 \left( {m_ -  R_T } \right)^2 \tau ^3 }} \left[ -\sin \left( {m_ -  \tau  - \frac{{\pi }}{4}} \right)^2 + \cos \left( {m_ -  \tau  - \frac{{\pi }}{4}} \right)^2 \right] + V _ - , \\
\rho _{wall}  &= \frac{{2m_ -  R_T ^2 \left( {\phi _ -   - \phi _T } \right)^2 }}{{\pi I_1 \left( {m_ -  R_T } \right)^2 \tau ^3 }}\cos \left( {m_ -  \tau  - \frac{{\pi }}{4}} \right)^2.
\end{align}
As expected from virial theorem, the average of kinetic and potential energy terms in the Lagrangian cancel out, because of the quadratic form of the potential around the true vacuum. However this kind of ideas can in general provide a cosmological constant that depends on the size of the bubble. It could be a promising candidate for the explanation of the small size of the observed cosmological constant.

\subsection{The Decay Rate}

The volcano potential may be used to approximate decay rates of metastable vacua, in the case of a potential that does not fit the requirements for the thin wall approximation. Thus, it is interesting to calculate the decay rate. The decay rate per unit volume is given by
\begin{equation}
\frac{\Gamma }{V} = Ae^{ - \frac{B}{\hbar }} \left[ {1 + \mathcal{O} \left( \hbar  \right)} \right],
\end{equation}
where coefficient $B$ is given by the Euclidean action of the tunneling solution
\begin{equation}
B = S_E \left[ {\phi \left( \rho  \right)} \right] - S_E \left[ {\phi _ +  } \right].
\end{equation}
Spherical symmetry of the tunneling solution implies
\begin{equation}
S_E \left[ {\phi \left( \rho  \right)} \right] = 2\pi ^2 \int_0^\infty  {\delta \rho \rho ^3 \left( {\frac{1}{2}\dot \phi \left( \rho  \right)^2  + V\left[ {\phi \left( \rho  \right)} \right]} \right)}.
\end{equation}
Using the formula of our solution we get
\begin{multline}
B = S_E \left[ {\phi \left( \rho  \right)} \right] - S_E \left[ {\phi _ +  } \right] = 2\pi ^2 \left( {V_ -   - V_ +  } \right)\int_0^{R_T } {\delta \rho \rho ^3 }\\
+ \frac{{\pi ^2 m_ -  ^2 R_T ^2 \left( {\phi _ -   - \phi _T } \right)^2 }}{{I_1 \left( {m_ -  R_T } \right)^2 }}\int_0^{R_T } {\delta \rho \rho \left[ {I_1 \left( {m_ -  \rho } \right)^2  + I_2 \left( {m_ -  \rho } \right)^2 } \right]}\\
 + \frac{{\pi ^2 m_ +  ^2 R_T ^2 \left( {\phi _T  - \phi _ +  } \right)^2 }}{{K_1 \left( {m_ +  R_T } \right)^2 }}\int_{R_T }^\infty  {\delta \rho \rho \left[ {K_1 \left( {m_ -  \rho } \right)^2  + K_2 \left( {m_ -  \rho } \right)^2 } \right]}.
\end{multline}
Applying properties of modified Bessel functions we find
\begin{multline}
 B = \frac{{\pi ^2 \left( {V_ -   - V_ +  } \right)R_T ^4 }}{2} \\
 + \pi ^2 m_ -  R_T ^3 \left( {\phi _ -   - \phi _T } \right)^2 \frac{{I_2 \left( {m_ -  R_T } \right)}}{{I_1 \left( {m_ -  R_T } \right)}} + \pi ^2 m_ +  R_T ^3 \left( {\phi _T  - \phi _ +  } \right)^2 \frac{{K_2 \left( {m_ +  R_T } \right)}}{{K_1 \left( {m_ +  R_T } \right)}}.
\end{multline}
Finally using equation (\ref{eq:volcnum}) we find
\begin{equation}
B = \frac{{\pi ^2 \left( {V_ -   - V_ +  } \right)R_T ^4 }}{2} + \pi ^2 m_ +  R_T ^3 \left( {\phi _ -   - \phi _ +  } \right)\left( {\phi _T  - \phi _ +  } \right)\frac{{K_2 \left( {m_ +  R_T } \right)}}{{K_1 \left( {m_ +  R_T } \right)}}.
\label{eq:volcdecay}
\end{equation}
The first term corresponds to the vacuum energy gained in the volume of the bubble, while the second term corresponds to the energy spent on the bubble wall.

It would be interesting to study whether the decay rate we just calculated reduces to the known formula in the thin wall limit. In the limit where the energy difference between the true and false vacuum is small, we can derive from the form of the potential that
\begin{equation}
\phi _T  = \frac{{m_ -  \phi _ -   + m_ +  \phi _ +  }}{{m_ -   + m_ +  }} + \frac{\varepsilon }{{m_ -  m_ +  \left( {\phi _ -   - \phi _ +  } \right)}} + \mathcal{O}\left( {\varepsilon ^2 } \right),
\end{equation}
where $\varepsilon \equiv V_+ - V_-$. We can use the above to find
\begin{equation}
\frac{{m_ +  }}{{m_ -  }}\frac{{\phi _T  - \phi _ +  }}{{\phi _ -   - \phi _T }} = 1 - \frac{\varepsilon }{{\mu ^2 \left( {\phi _ -   - \phi _ +  } \right)^2 }} + \mathcal{O}\left( {\varepsilon ^2 } \right).
\end{equation}
That means that the right hand side of equation (\ref{eq:volcnum}) is very close to one, thus $R_T$ has to be large, as expected for the thin wall limit. This allows us to use asymptotic formulas for the modified Bessel functions to approximate the left hand side of (\ref{eq:volcnum}) as
\begin{equation}
\frac{{K_1 \left( {m_ +  R_T } \right)}}{{K_2 \left( {m_ +  R_T } \right)}}\frac{{I_2 \left( {m_ +  R_T } \right)}}{{I_1 \left( {m_ +  R_T } \right)}} = 1 - \frac{3}{{2\mu R_T }} + \mathcal{O}\left( {\frac{1}{{R_T ^2 }}} \right),
\end{equation}
where
\begin{equation}
\mu  \equiv \frac{{m_ -  m_ +  }}{{m_ -   + m_ +  }}.
\end{equation}
This allows us to calculate the size of the emitted bubble to be
\begin{equation}
R_T  = \frac{{3\mu \left( {\phi _ -   - \phi _ +  } \right)^2 }}{{2\varepsilon }} + \mathcal{O}\left( {\varepsilon ^0 } \right)
\end{equation}
and finally substituting to formula (\ref{eq:volcdecay}) we result in
\begin{equation}
B = \frac{{27\pi ^2 \mu ^4 \left( {\phi _ -   - \phi _ +  } \right)^8 }}{{32\varepsilon ^3 }} + \mathcal{O}\left( {\frac{1}{{\varepsilon ^2 }}} \right).
\label{eq:volcthinwall}
\end{equation}

In thin wall approximation as described in \cite{Coleman:1977py}, the $B$ factor equals
\begin{equation}
B = \frac{{27\pi ^2 S_1 ^4 }}{{2\varepsilon ^3 }},
\label{eq:thinwall}
\end{equation}
where
\begin{equation}
S_1  = \int_{\phi _ +  }^{\phi _ -  } {d\phi \sqrt {2V\left( \phi  \right) - V_ -  } }.
\end{equation}
In our case it is not difficult to calculate $S_1$, when the vacua energies are close
\begin{equation}
S_1  = \int_{\phi _ +  }^{\frac{{m_ -  \phi _ -   + m_ +  \phi _ +  }}{{m_ -   + m_ +  }}} {d\phi m_ +  \left( {\phi  - \phi _ +  } \right)}  + \int_{\frac{{m_ -  \phi _ -   + m_ +  \phi _ +  }}{{m_ -   + m_ +  }}}^{\phi _ -  } {d\phi m_ -  \left( {\phi _ -   - \phi } \right)}  + \mathcal{O}\left( \varepsilon  \right)
\end{equation}
and after some algebra
\begin{equation}
S_1  = \frac{{\mu \left( {\phi _ -   - \phi _ +  } \right)^2 }}{2} + \mathcal{O}\left( \varepsilon  \right).
\end{equation}
Substituting the latter to (\ref{eq:thinwall}) gives us exactly the same result as (\ref{eq:volcthinwall}).

Another interesting limit to check is the tunneling without barrier limit, which is discussed in \cite{Lee:1985uv}. In our case this clearly corresponds to the limit $m_ +   \to 0$. In this limit, we can use asymptotic formulas for the modified Bessel functions of the second kind to approximate
\begin{equation}
\frac{{K_1 \left( {m_ -  R_T } \right)}}{{K_2 \left( {m_ -  R_T } \right)}} = \frac{{m_ -  R_T }}{2} + \mathcal{O}\left( {m_ -  ^2 } \right).
\end{equation}
This lets us write equation (\ref{eq:volcnum}) as
\begin{equation}
\frac{{m_ -  R_T }}{2}\frac{{I_2 \left( {m_ -  R_T } \right)}}{{I_1 \left( {m_ -  R_T } \right)}} = \frac{{\phi _T  - \phi _ +  }}{{\phi _ -   - \phi _T }}.
\end{equation}
The same asymptotic expansions allow us to write equation (\ref{eq:volcdecay}) as
\begin{equation}
B = \frac{{\pi ^2 \left( {V_ -   - V_ +  } \right)R_T ^4 }}{2} + 2\pi ^2 R_T ^2 \left( {\phi _ -   - \phi _ +  } \right)\left( {\phi _T  - \phi _ +  } \right).
\end{equation}

Now we distinguish two cases. If $\phi _T  - \phi _ + \ll \phi _ -   - \phi _T$ we can use asymptotic expansions of modified bessel functions for small arguments and equation (\ref{eq:volcnum}) can be written as
\begin{equation}
\frac{{\left( {m_ -  R_T } \right)^2 }}{8} + \frac{{\left( {m_ -  R_T } \right)^4 }}{{192}} + \mathcal{O}\left( {m_ -  ^6 R_T ^6 } \right) = \frac{{\phi _T  - \phi _ +  }}{{\phi _ -   - \phi _T }}.
\end{equation}
We can solve for $R_T$
\begin{equation}
R_T ^2  = \frac{8}{{m_ -  ^2 }}\frac{{\phi _T  - \phi _ +  }}{{\phi _ -   - \phi _T }} + \frac{8}{{3m_ -  ^2 }}\left( {\frac{{\phi _T  - \phi _ +  }}{{\phi _ -   - \phi _T }}} \right)^2  + \mathcal{O}\left[ {\left( {\frac{{\phi _T  - \phi _ +  }}{{\phi _ -   - \phi _T }}} \right)^3 } \right].
\end{equation}
Substituting in (\ref{eq:volcdecay}) and using $V_ -   - V_ +   = \frac{1}{2}m_ -  ^2 \left( {\phi _ -   - \phi _T } \right)^2$ we find
\begin{equation}
B = \frac{{16\pi ^2 \left( {\phi _T  - \phi _ +  } \right)^3 \left( {\phi _ -   - \phi _T } \right)}}{{3\left( {V_ -   - V_ +  } \right)}} + \mathcal{O}\left[ {\left( {\phi _T  - \phi _ +  } \right)^4 } \right].
\end{equation}
This agrees with the results in \cite{Lee:1985uv} up to a factor of 2 that occurs because in our case the rolling region of the potential is quadratic instead of linear.

If $\phi _T  - \phi _ + \gg \phi _ -   - \phi _T$ we can use asymptotic expansions of modified bessel functions for large arguments and equation (\ref{eq:volcnum}) can be written as
\begin{equation}
\frac{{m_ -  R_T }}{2} + \mathcal{O}\left( {m_ -  ^0 R_T ^0 } \right) = \frac{{\phi _T  - \phi _ +  }}{{\phi _ -   - \phi _T }}.
\end{equation}
Again we solve for $R_T$
\begin{equation}
R_T ^2  = \frac{4}{{m_ -  ^2 }}\left( {\frac{{\phi _T  - \phi _ +  }}{{\phi _ -   - \phi _T }}} \right)^2  + \mathcal{O}\left[ {\left( {\frac{{\phi _T  - \phi _ +  }}{{\phi _ -   - \phi _T }}} \right)^0 } \right]
\end{equation}
and substitute in (\ref{eq:volcdecay}) to find
\begin{equation}
B = \frac{{2\pi ^2 \left( {\phi _T  - \phi _ +  } \right)^4 }}{{\left( {V_ -   - V_ +  } \right)}} + \mathcal{O}\left[ {\left( {\phi _T  - \phi _ +  } \right)^2 } \right],
\end{equation}
which agrees with the results of \cite{Lee:1985uv}.

\section{A Smooth Quadratic Potential and the Size of the Emitted Bubble}

\subsection{The Approximation}

\begin{figure}[h]
\begin{center}
\includegraphics[angle=0,width=0.65\textwidth]{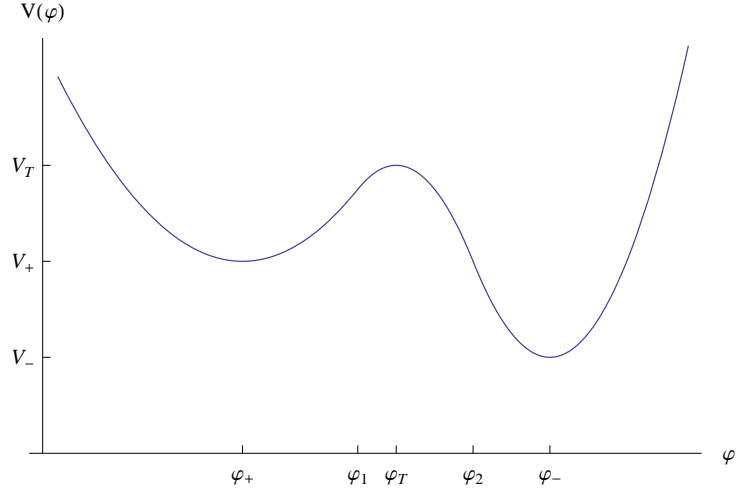}
\end{center}
\caption{The smooth quadratic approximation}
\label{fig:smoothpot}
\end{figure}

The volcano approximation makes a bad non-smooth description of the barrier top. We can use the fact that quadratic potentials are solvable, in order to improve our approximation, and search for new qualitative properties of the solutions that originate from the form of the potential at the barrier top. An appropriate approximation is
\begin{equation}
 V\left( \phi  \right) =
 \begin{cases}
 \cfrac{1}{2}m_ +  ^2 \left( {\phi  - \phi _ +  } \right)^2  + V_ + , & \phi < \phi_1 \\
 -\cfrac{1}{2}m_ T  ^2 \left( {\phi  - \phi _ T  } \right)^2  + V_ T , & \phi_1 < \phi < \phi_2 \\
 \cfrac{1}{2}m_ -  ^2 \left( {\phi  - \phi _ -  } \right)^2  + V_ - , & \phi > \phi_2 \\
 \end{cases},
\end{equation}
which is plotted in figure \ref{fig:smoothpot}.
Such an approximation provides quite a large flexibility in fitting an arbitrary potential. We can select the above potential in such a way that matches the actual potential in positions and energies of the true and false vacuum, as well as the top of barrier, and moreover the curvature of the potential in these positions or alternatively select so that it matches the positions and energies of the vacua, as well as one the three aforementioned curvatures and simultaneously be smooth at $\phi_1$ and $\phi_2$.

\subsection{The Instanton Solution}

First we note that at the region between $\phi_1$ and $\phi_2$ the equation of motion can be solved in exactly the same way as we did in the volcano potential, with the only deference of getting the Bessel functions instead of the modified Bessel function. The equation of motion is
\begin{equation}
\ddot \phi  + \frac{3}{\rho }\dot \phi  = m^2 \left( {\phi  - \phi _0 } \right),
\end{equation}
and the general solution can be written as
\begin{equation}
\phi  = \phi _0  + \frac{{c_1 J_1 \left( {m\rho } \right) + c_2 Y_1 \left( {m\rho } \right)}}{\rho }.
\end{equation}
For later use in the matching condition we also calculate the derivative of the solution
\begin{equation}
\dot \phi  =  - m \frac{{c_1 J_2 \left( {m\rho } \right) + c_2 K_2 \left( {m\rho } \right)}}{\rho }.
\end{equation}

Similarly to the volcano potential, we expect that there are no solutions that reach the true vacuum in Euclidean space. We separate two cases. In the first the solution does not reach $\phi_2$ inside Euclidean space and in the second it does. Here we will study the first case, as it is simpler and provides the information we want. Such a solution will look like
\begin{equation}
 \phi  =
 \begin{cases}
 \phi _T  + \cfrac{{c_{1T} J_1 \left( {m_T \rho } \right) + c_{2T} Y_1 \left( {m_T \rho } \right)}}{\rho } , & \rho < R_1 \\
 \phi _ +   + \cfrac{{c_{1 + } I_1 \left( {m_ +  \rho } \right) + c_{2 + } K_1 \left( {m_ +  \rho } \right)}}{\rho } , & R_1 < \rho < R_+ \\
 \phi_+ , \phantom{\cfrac{{1}}{{2}}} & \rho > R_+ \\
 \end{cases}.
\end{equation}
Exactly as in the volcano potential case, $R_+$ has to be infinite, and the boundary conditions for the solution imply
\begin{align}
c_{2T}  &= 0, \\
c_{1 + }  &= 0.
\end{align}
Demanding that $\mathop {\lim }\limits_{\rho \to {R_1}^- } \phi \left( \rho \right) = \mathop {\lim }\limits_{\rho \to {R_1}^+ } \phi \left( \rho \right) = \phi _ 1$ gives us
\begin{align}
c_{1T}  &= \frac{{\phi _1  - \phi _T }}{{J_1 \left( {m_T R_1 } \right)}}R_1 , \\
c_{2 + }  &= \frac{{\phi _1  - \phi _ +  }}{{K_1 \left( {m_ +  R_1 } \right)}}R_1.
\end{align}
Finally smoothness at $\rho=R_1$ implies
\begin{equation}
\frac{{K_1 \left( {m_ +  R_1 } \right)}}{{K_2 \left( {m_ +  R_1 } \right)}}\frac{{J_2 \left( {m_T R_1 } \right)}}{{J_1 \left( {m_T R_1 } \right)}} =  - \frac{{m_ +  }}{{m_T }}\frac{{\phi _1  - \phi _ +  }}{{\phi _T  - \phi _1 }}.
\label{eq:smoothnum}
\end{equation}
Thus the instanton solution is given by
\begin{equation}
 \phi  =
 \begin{cases}
 \phi _T  - \cfrac{{R_1 \left( {\phi _T  - \phi _1 } \right)}}{\rho }\cfrac{{J_1 \left( {m_T \rho } \right)}}{{J_1 \left( {m_T R_1 } \right)}} , & \rho < R_1 \\
 \phi _ +   + \cfrac{{R_1 \left( {\phi _1  - \phi _ +  } \right)}}{\rho }\cfrac{{K_1 \left( {m_ +  \rho } \right)}}{{K_1 \left( {m_ +  R_1 } \right)}} , & \rho > R_1 \\
 \end{cases},
 \label{eq:smooth1solution}
\end{equation}
where $R_1$ is given by (\ref{eq:smoothnum}).

\subsection{Uniqueness of the solution and the Radius of the Emitted Bubble}

\begin{figure}[h]
\begin{center}
\includegraphics[angle=0,width=0.45\textwidth]{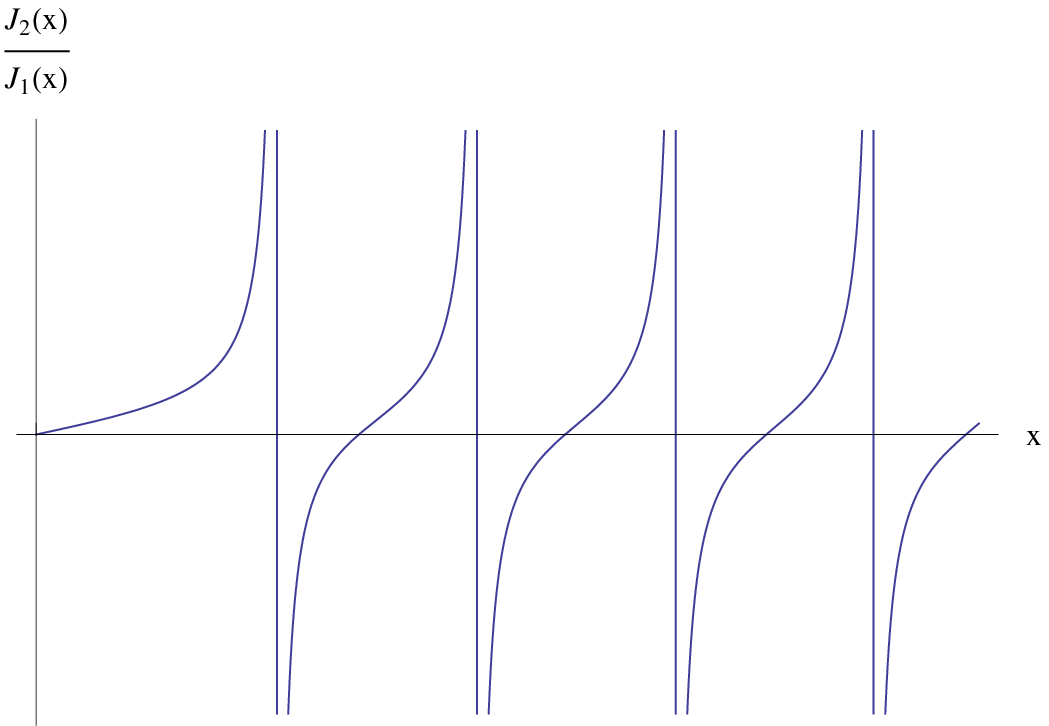}
\includegraphics[angle=0,width=0.45\textwidth]{volcanocondition2}
\end{center}
\caption{The numerical part of the solution}
\label{fig:smooth1condition}
\end{figure}

As in previous cases the existence or non-existence of a solution is decided by the last equation, that occurs by the demand of smoothness of the solution, in our case equation (\ref{eq:smoothnum}). In figure \ref{fig:smooth1condition} we plot $\frac{{K_1 \left( x \right)}}{{K_2 \left( x \right)}}$ and $\frac{{J_2 \left( x \right)}}{{J_1 \left( x \right)}}$. The first graph implies that actually we are going to have infinite solutions.

\begin{figure}[h]
\begin{center}
\includegraphics[angle=0,width=0.65\textwidth]{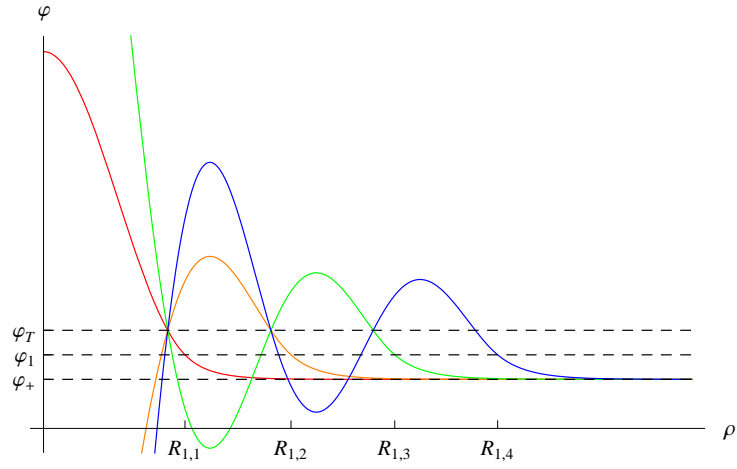}
\end{center}
\caption{The first solutions in the smooth quadratic approximation}
\label{fig:smooth1solutions}
\end{figure}

In figure \ref{fig:smooth1solutions} we plot the first solutions.
Actually any other solution except for the first one is not valid, since for some region of $\rho$ with $\rho<R_1$ region, it is true that $\phi<\phi_1$. Let' sketch a proof for that.

The function $\cfrac{{J_1 \left( {m_T \rho } \right)}}{{\rho}}$ describes an oscillation whose amplitude decreases monotonically, as it can be seen in figure \ref{fig:smooth1solutions}. Based on that and using properties of Bessel functions, one can show that if $\rho_S$ is a stationary point of this function, then
\begin{equation}
\left| {\frac{{J_1 \left( {m_T \rho _S } \right)}}{{\rho _S }}} \right| > \frac{{J_1 \left( {m_T \rho } \right)}}{\rho },
\end{equation}
for any $\rho > \rho_S$. Applying this for $\rho=R_1$ in equation \ref{eq:smooth1solution} it is easy to find that for any stationary point $\rho_S$, with $\rho_S<R_1$ it is true that
\begin{equation}
\left| {\phi _S  - \phi _T } \right| > \phi _T  - \phi _1.
\end{equation}
For any solution except for the first one there are at least two stationary points in the region $\rho<R_1$. For at least one of them $\phi_S$ is going to be smaller than $\phi_T$, thus making the above equation
\begin{equation}
\phi_S<\phi_1.
\end{equation}
This means that the solution for $\rho<R_1$ has entered a region of values for which it is not valid. Thus any solution except for the first one is not valid, the solution is unique.

This is not a general proof of the uniqueness of the solution for a general potential, however in this case the oscillatory behavior of the solution in the region of the barrier top resembles exactly the general behavior that could produce multiple solutions in general.

So the only solution it may hold is the first solution, which is valid if $\phi \left( 0 \right) < \phi_2$, or else
\begin{equation}
\frac{{2J_1 \left( {m_T R_1 } \right)}}{{m_T R_1 }} <  - \frac{{\phi _T  - \phi _1 }}{{\phi _2  - \phi _T }}.
\end{equation}
If this condition does not hold we should search for a solution of the second case.

Figure \ref{fig:smooth1condition} implies that the curvature of the potential at the top is strongly related with the size of the emitted bubble. As we stated above, the only valid solution is the one corresponding to the smallest solution of equation (\ref{eq:smoothnum}). From figure \ref{fig:smooth1condition} we can see, that this solution always satisfies
\begin{equation}
\frac{{\rho_{1,1}}}{{m_T}} < R _ 1 < \frac{{\rho_{2,1}}}{{m_T}},
\end{equation}
where $\rho_{\alpha,n}$ is the n'th root of $J_\alpha$. We note that $\rho_{1,1}\simeq 3.83$ and $\rho_{2,1}\simeq 5.14$.
So the characteristic radius of the emitted bubble is always of the order $\frac{{1}}{{m_T}}$.

As this case of solutions is the one that actually lies as far as possible from the thin wall approximation, in which the radius of the bubble tends to infinity, we expect that actually $\frac{{\rho_{1,1}}}{{m_T}}$ serves as a general lower bound for the radius of the emitted bubble.

\subsection{The Decay Rate}

As we did in the volcanic approximation, we can calculate the $B$ factor of decay rate from the Euclidean action. Using the form of our solution we find
\begin{multline}
B = S_E \left[ {\phi \left( \rho  \right)} \right] - S_E \left[ {\phi _ +  } \right] = 2\pi ^2 \left( {V_ T   - V_ +  } \right)\int_0^{R_T } {\delta \rho \rho ^3 }\\
+ \frac{{\pi ^2 m_ T  ^2 R_1 ^2 \left( {\phi _ T   - \phi _1 } \right)^2 }}{{J_1 \left( {m_ T  R_1 } \right)^2 }}\int_0^{R_1 } {\delta \rho \rho \left[ {-J_1 \left( {m_ T  \rho } \right)^2  + J_2 \left( {m_ T  \rho } \right)^2 } \right]}\\
 + \frac{{\pi ^2 m_ +  ^2 R_1 ^2 \left( {\phi _1  - \phi _ +  } \right)^2 }}{{K_1 \left( {m_ +  R_1 } \right)^2 }}\int_{R_1 }^\infty  {\delta \rho \rho \left[ {K_1 \left( {m_ -  \rho } \right)^2  + K_2 \left( {m_ -  \rho } \right)^2 } \right]}.
\end{multline}
Applying properties of Bessel functions and modified Bessel functions we find
\begin{multline}
B = \frac{{\pi ^2 \left( {V_T  - V_ +  } \right)R_1 ^4 }}{2}\\
 - \pi ^2 m_T R_1 ^3 \left( {\phi _T  - \phi _1 } \right)^2 \frac{{J_2 \left( {m_T R_1 } \right)}}{{J_1 \left( {m_T R_1 } \right)}} + \pi ^2 m_ +  R_1 ^3 \left( {\phi _1  - \phi _ +  } \right)^2 \frac{{K_2 \left( {m_ +  R_1 } \right)}}{{K_1 \left( {m_ +  R_1 } \right)}}.
\end{multline}
Finally using equation (\ref{eq:smoothnum}) we find
\begin{equation}
B = \frac{{\pi ^2 \left( {V_T  - V_ +  } \right)R_1 ^4 }}{2} + \pi ^2 m_ +  R_1 ^3 \left( {\phi _T  - \phi _ +  } \right)\left( {\phi _1  - \phi _ +  } \right)\frac{{K_2 \left( {m_ +  R_1 } \right)}}{{K_1 \left( {m_ +  R_1 } \right)}}.
\end{equation}

\subsection{The Second Case of Solutions on the Smooth Quadratic Potential}

\begin{figure}[h]
\begin{center}
\includegraphics[angle=0,width=0.65\textwidth]{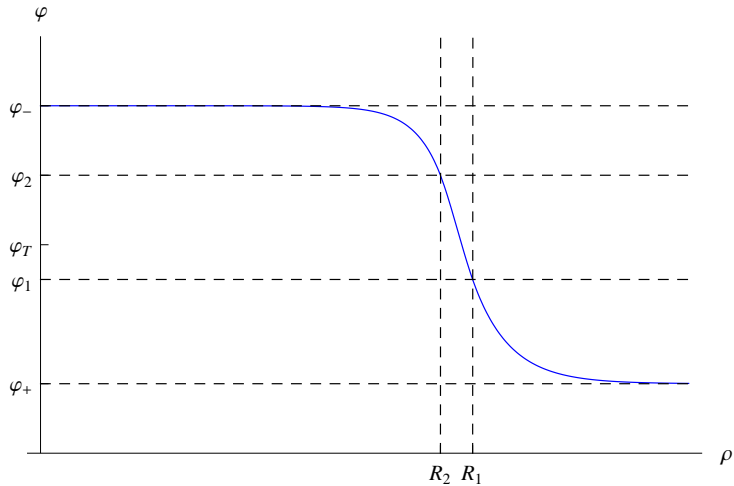}
\end{center}
\caption{The solution for the smooth quadratic potential}
\label{fig:smooth2solution}
\end{figure}

An instanton belonging in the second case will be of the form
\begin{equation}
 \phi  =
 \begin{cases}
 \phi _ -   + \cfrac{{c_{1 - } I_1 \left( {m_ -  \rho } \right)}}{\rho } , & \rho < R_2 \\
 \phi _T  + \cfrac{{c_{1T} J_1 \left( {m_T \rho } \right) + c_{2T} Y_1 \left( {m_T \rho } \right)}}{\rho } , & R_2 < \rho < R_1 \\
 \phi _ +   + \cfrac{{c_{2 + } K_1 \left( {m_ +  \rho } \right)}}{\rho } , & \rho > R_1 \\
 \end{cases},
\end{equation}
where we have already embodied the necessary boundary conditions. Demanding that $\phi \left( R_1 \right) = \phi_1$ and $\phi \left( R_2 \right) = \phi_2$ and continuity results in
\begin{align}
c_{1 - }  &= \frac{{\phi _2  - \phi _ -  }}{{I_1 \left( {m_ -  R_2 } \right)}}R_2, \\
c_{2 + }  &= \frac{{\phi _1  - \phi _ +  }}{{K_1 \left( {m_ +  R_1 } \right)}}R_1
\end{align}
and
\begin{align}
c_{1T}  &= - \frac{{Y_1 \left( {m_T R_2 } \right)\left( {\phi _T  - \phi _1 } \right)R_1  - Y_1 \left( {m_T R_1 } \right)\left( {\phi _2  - \phi _T } \right)R_2 }}{{J_1 \left( {m_T R_1 } \right)Y_1 \left( {m_T R_2 } \right) - J_1 \left( {m_T R_2 } \right)Y_1 \left( {m_T R_1 } \right)}},\\
c_{2T}  &= - \frac{{J_1 \left( {m_T R_1 } \right)\left( {\phi _2  - \phi _T } \right)R_2  - J_1 \left( {m_T R_2 } \right)\left( {\phi _T  - \phi _1 } \right)R_1 }}{{J_1 \left( {m_T R_1 } \right)Y_1 \left( {m_T R_2 } \right) - J_1 \left( {m_T R_2 } \right)Y_1 \left( {m_T R_1 } \right)}}.
\end{align}
Finally demanding smoothness results in the following set of equations
\begin{align}
m_T \left[ {c_{1T} J_2 \left( {m_T R_1 } \right) + c_{2T} Y_2 \left( {m_T R_1 } \right)} \right] &=  - m_ -  c_{1 - } I_2 \left( {m_ -  R_2 } \right),\\
m_T \left[ {c_{1T} J_2 \left( {m_T R_2 } \right) + c_{2T} Y_2 \left( {m_T R_2 } \right)} \right] &= m_ +  c_{2 + } K_2 \left( {m_ +  R_1 } \right).
\end{align}
These two equations provide a solution for $R_1$ and $R_2$. Unfortunately the problem is very complicated, and has to be numerically solved. In figure \ref{fig:smooth2solution} we show how a solution to this problem looks like.

\section{A Triangular Potential and a Candidate for Dark Energy}
\label{sec:triangular}

\subsection{The Approximation}

\begin{figure}[h]
\begin{center}
\includegraphics[angle=0,width=0.65\textwidth]{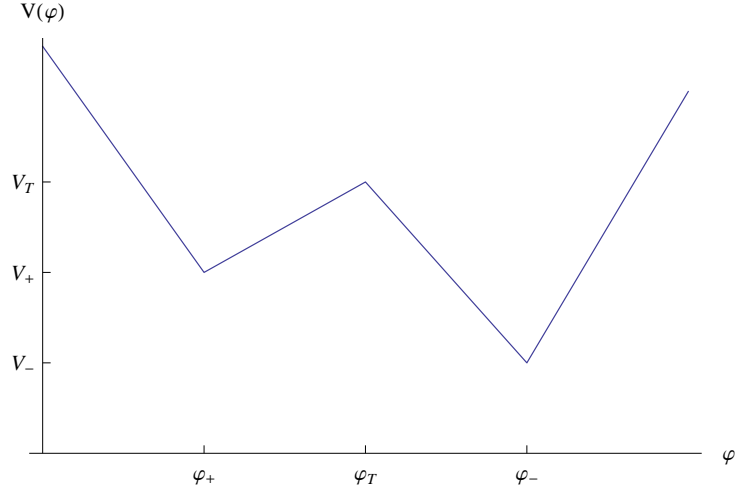}
\end{center}
\caption{The triangular approximation}
\label{fig:trigpot}
\end{figure}

We will now study the case of a potential barrier approximated by a segment of linear potentials. Such a potential looks like in figure \ref{fig:trigpot}. This approximation has been analyzed in \cite{Duncan:1992ai}. We review this derivation and then analytically continue to negative proper time.

In the following we use the definitions
\begin{equation}
\Delta V_ \pm   \equiv V_T  - V_ \pm  ,\quad \Delta \phi _ \pm   \equiv  \pm \left( {\phi _T  - \phi _ \pm  } \right),
\end{equation}
\begin{equation}
\lambda _ \pm   \equiv \frac{{\Delta V_ \pm  }}{{\Delta \phi _ \pm  }},\quad c \equiv \frac{{\lambda _ -  }}{{\lambda _ +  }}.
\end{equation}
In other words the potential barrier is described by
\begin{equation}
 V\left( \phi  \right) =
 \begin{cases}
 \lambda _ +  \left( {\phi  - \phi _ +  } \right) + V_ + , & \phi_+ < \phi < \phi_T \\
 - \lambda _ -  \left( {\phi  - \phi _ -  } \right) + V_ - , & \phi_T < \phi < \phi_- \\
 \end{cases}
\end{equation}

\subsection{The Instanton Solution}

In order to find the appropriate solution we need to solve the equation
\begin{equation}
\ddot \phi  + \frac{3}{\rho}\dot \phi  = \lambda.
\end{equation}
It is not difficult to show that the general solution is
\begin{equation}
\phi  = \frac{\lambda }{8}\rho^2  + \frac{{c}}{{\rho^2 }} + K.
\end{equation}

Using the above, an instanton that does not reach the true vacuum in Euclidean space will look like
\begin{equation}
 \phi  =
 \begin{cases}
 - \cfrac{{\lambda _ -  }}{8}\rho^2  + \cfrac{{c_ -  }}{{\rho^2 }} + K_ - , & \rho < R_T \\
 \cfrac{{\lambda _ +  }}{8}\rho^2  + \cfrac{{c_ +  }}{{\rho^2 }} + K_ + , & R_T < \rho < R_+ \\
 \phi_+ , \phantom{\cfrac{{1}}{{2}}} & \rho > R_+ \\
 \end{cases}.
\end{equation}

Let's now apply the appropriate boundary and matching conditions to determine the constants as well as the radii $R_T$ and $R_+$.
The field must be stationary at the origin. This implies that
\begin{equation}
c_-  = 0.
\end{equation}
Demanding continuity of the derivative of the field at $R=R_+$ gives us
\begin{equation}
c_+  = \frac{{\lambda _ +  }}{8}R_ +  ^4.
\end{equation}
Then we demand continuity at $R=R_+$ and $R=R_T$ and we find
\begin{equation}
K_ +   = \phi _ +   - \frac{{\lambda _ +  }}{4}R_ +  ^2,
\end{equation}
\begin{equation}
K_ -   = \phi _ +   + \frac{{\lambda _ +  }}{{8R_T ^2 }}\left( {R_ +  ^2  - R_T ^2 } \right)^2  + \frac{{\lambda _ -  }}{8}R_T ^2  \equiv \phi _0 = \phi _ T  + \frac{{\lambda _ -  }}{8}R_T ^2  \equiv \phi _0.
\label{eq:trigphinot}
\end{equation}
Continuity of the derivative at $R=R_T$ gives us
\begin{equation}
R_ +  ^4  = \left( {1 + c} \right)R_T ^4.
\label{eq:trigrplus}
\end{equation}
So far we have expressed all unknowns in terms of the unknown radius $R_T$. In order to determine this final unknown we demand that $\phi \left( R_T \right) = \phi_T$. We result in
\begin{equation}
\phi _T  - \phi _ +   = \frac{{\lambda _ +  }}{8}\left( {\sqrt {1 + c}  - 1} \right)^2 R_T ^2.
\label{eq:trigrt}
\end{equation}

Now we have completely determine the tunneling solution. To sum up
\begin{equation}
 \phi  =
 \begin{cases}
 \phi _0  - \cfrac{{\lambda _ -  }}{8}\rho^2 , & \rho < R_T \\
 \phi _ +   + \cfrac{{\lambda _ +  }}{8}\cfrac{1}{{\rho ^2 }}\left( {\rho ^2  - R_ +  ^2 } \right)^2 , & R_T < \rho < R_+ \\
 \phi_+ , \phantom{\cfrac{{1}}{{2}}} & \rho > R_+ \\
 \end{cases},
 \label{eq:trigsolution}
\end{equation}
where the two radii are given by (\ref{eq:trigrplus}) and (\ref{eq:trigrt}) and the value of the field at the origin $\phi_0$ is given by (\ref{eq:trigphinot}).

If we want our solution to make sense it has to be that $\phi_0<\phi_-$. Otherwise the field has already reached the true vacuum in Euclidean space. The above condition implies that
\begin{equation}
\frac{{\phi _ -   - \phi _T }}{{\phi _T  - \phi _ +  }} > \frac{c}{{\left( {\sqrt {1 + c}  - 1} \right)^2 }}.
\label{eq:trigcondition}
\end{equation}
If this condition holds the field never reaches the true vacuum in Euclidean space and thus we expect to perform a damped oscillation around the true vacuum in the interior of the bubble. Otherwise the field equals exactly to $\phi_-$ inside a sphere of finite radius in Euclidean space, thus the analytical continuation for imaginary Euclidean time is trivially $\phi=\phi_-$. Such a solution is well analyzed in \cite{Duncan:1992ai} and we will not study it here.

\subsection{The Analytical Continuation to Lorentzian Spacetime}

We are interested in studying the solution for imaginary Euclidean radius. As we commended in section \ref{subsec:kinds}, this is interesting only if condition (\ref{eq:trigcondition}) holds. In such case the solution is given by (\ref{eq:trigsolution}). We substitute $\rho=i \tau$ in the formula for the solution that is valid for $\rho<R_T$ to get
\begin{equation}
\phi _0  + \frac{{\lambda _ -  }}{8}\tau ^2.
\label{eq:trigsolin0}
\end{equation}
This grows indefinitely as $\tau$ decreases. This means that at some finite $\tau$ it reaches the true vacuum. After that point, the solution we have is not valid anymore and we need to find an appropriate solution for $\phi>\phi_-$. We approximate the potential around the true vacuum as
\begin{equation}
 V\left( \phi  \right) =
 \begin{cases}
 - \lambda ^ -  \left( {\phi  - \phi _ -  } \right) + V_ - , & \phi < \phi_- \\
 \lambda ^ +  \left( {\phi  - \phi _ -  } \right) + V_ - , & \phi > \phi_- \\
 \end{cases},
\end{equation}
where obviously $\lambda^-=\lambda_-$. So after the $\tau$ where the solution reaches the true vacuum, we have to fit a solution of the form
\begin{equation}
\phi  =  - \frac{{\lambda ^ +  }}{8}\tau ^2  + \frac{c}{{\tau ^2 }} + K.
\label{eq:trigsolin1}
\end{equation}
However this solution will reach a maximum and return to the true vacuum, with some non-vanishing derivative, thus enforcing us to fit again a solution of the form
\begin{equation}
\phi  = \frac{{\lambda ^ -  }}{8}\tau ^2  + \frac{c}{{\tau ^2 }} + K,
\label{eq:trigsolin2}
\end{equation}
and so on. So we are going to get an infinite sequence of segments describing a damped oscillation around the true vacuum. So we need to solve the general problem of fitting a solution of the form (\ref{eq:trigsolin1}) or (\ref{eq:trigsolin2}) to the boundary conditions
\begin{equation}
\phi \left( {T_0 } \right) = \phi _ -  ,\quad \dot \phi \left( {T_0 } \right) = \dot \Phi _ 0,
\end{equation}
or else solve the matching conditions
\begin{align}
\phi _ -   &=  \pm \cfrac{{\lambda ^ \mp  }}{8}T_0 ^2  + \cfrac{c}{{T_0 ^2 }} + K, \\
\dot \Phi _ 0  &=  \pm \cfrac{{\lambda ^ \mp  }}{4}T_0  - \cfrac{{2c}}{{T_0 ^3 }}.
\end{align}
It is not difficult to find that the solution is
\begin{align}
 c &=  \pm \cfrac{{\lambda ^ \mp  }}{8}T_0 ^4  - \cfrac{{\dot \Phi _ 0 T_0 ^3 }}{2}, \\
 K &= \phi _ -   \mp \cfrac{{\lambda ^ \mp  }}{4}T_0 ^2  + \cfrac{{\dot \Phi _ 0 T_0 }}{2}.
\end{align}
Once we found the right expression we need to find the new point where the solution reaches the true vacuum. Demanding that $\phi \left( {T_1 } \right) = \phi _ -$, we take
\begin{align}
 T_1 ^2  &= T_0 ^2  \mp \cfrac{{4\dot \Phi _0 T_0 }}{{\lambda ^ \mp  }}, \\
 \dot \phi \left( {T_1 } \right) \equiv \dot \Phi _1  &=  - \dot \Phi _0 \frac{{T_0 }}{{T_1 }}.
\end{align}
Thus we can express the solution inductively as
\begin{equation}
 \phi =
 \begin{cases}
 - \cfrac{{\lambda ^ +  }}{8}\tau ^2  + \cfrac{{c_{2n + 1} }}{{\tau ^2 }} + K_{2n + 1} ,& T_{2n + 1}  < \tau  < T_{2n} \\
 \cfrac{{\lambda ^ -  }}{8}\tau ^2  + \cfrac{{c_{2n} }}{{\tau ^2 }} + K_{2n} ,& T_{2n}  < \tau  < T_{2n - 1} \\
 \end{cases},
\end{equation}
where the constants in solution are given by
\begin{align}
c_{2n + 1}  &=  - \frac{{\lambda ^ +  }}{8}T_{2n} ^4  - \frac{{\dot \Phi _{2n} T_{2n} ^3 }}{2}, & c_{2n}  &= \frac{{\lambda ^ -  }}{8}T_{2n - 1} ^4  - \frac{{\dot \Phi _{2n - 1} T_{2n - 1} ^3 }}{2}, \\
K_{2n + 1}  &= \phi _ -   + \frac{{\lambda ^ +  }}{4}T_{2n} ^2  + \frac{{\dot \Phi _{2n} T_{2n} }}{2}, & K_{2n}  &= \phi _ -   - \frac{{\lambda ^ -  }}{4}T_{2n - 1} ^2  + \frac{{\dot \Phi _{2n - 1} T_{2n - 1} }}{2},
\end{align}
and the $T_n$'s and $\dot \Phi _{n}$'s can be calculated inductively by
\begin{align}
T_{2n} ^2  &= T_{2n - 1} ^2  + \frac{{4\dot \Phi _{2n - 1} T_{2n - 1} }}{{\lambda ^ +  }}, & T_{2n + 1} ^2  &= T_{2n} ^2  - \frac{{4\dot \Phi _{2n} T_{2n} }}{{\lambda ^ -  }}, \label{eq:trigcontrec1}\\
\dot \Phi _{2n}  &=  - \dot \Phi _{2n - 1} \frac{{T_{2n - 1} }}{{T_{2n} }}, & \dot \Phi _{2n+1} &=  - \dot \Phi _{2n} \frac{{T_{2n} }}{{T_{2n + 1} }}.
\label{eq:trigcontrec2}
\end{align}
\begin{figure}[h]
\begin{center}
\includegraphics[angle=0,width=0.65\textwidth]{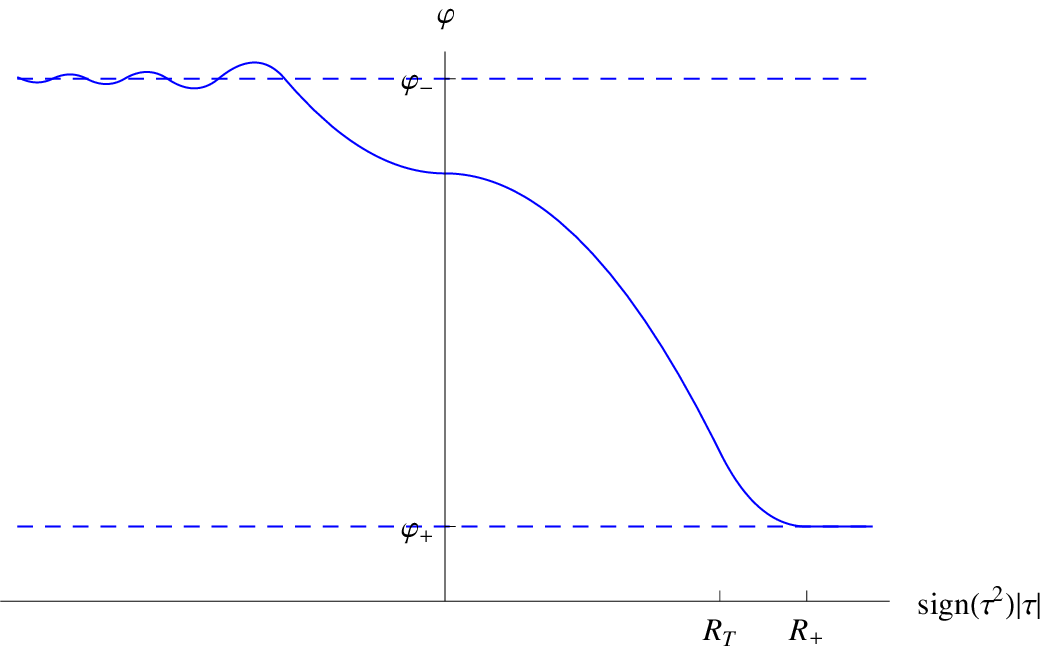}
\end{center}
\caption{The Instanton Solution}
\label{fig:trigsol}
\end{figure}
Finally the initial values for $T$ and $\dot \Phi$ can be easily calculated by equation (\ref{eq:trigsolin0}) (which also defines $c_0$ and $K_0$)
\begin{align}
T_0 ^2  &= \frac{8}{{\lambda ^ -  }}\left( {\phi _ -   - \phi _0 } \right) \\
\dot \Phi _0  &= \frac{{\lambda ^ -  }}{4}T_0.
\end{align}
The solution is plotted in figure \ref{fig:trigsol}.

\subsection{Asymptotic Behavior of the Solution}

For large $\tau$ expressions (\ref{eq:trigcontrec1}) and (\ref{eq:trigcontrec2}) can be approximated by
\begin{align}
T_{2n + 1}  &\simeq T_{2n}  - \frac{{2\dot \Phi _{2n} }}{{\lambda ^ -  }}, & T_{2n}  &\simeq T_{2n - 1}  + \frac{{2\dot \Phi _{2n - 1} }}{{\lambda ^ +  }},\\
\dot \Phi _{2n + 1}  &\simeq  - \dot \Phi _{2n} \left( {1 + \frac{{2\dot \Phi _{2n} }}{{\lambda ^ -  T_{2n} }}} \right), & \dot \Phi _{2n}  &\simeq  - \dot \Phi _{2n - 1} \left( {1 - \frac{{2\dot \Phi _{2n - 1} }}{{\lambda ^ +  T_{2n - 1} }}} \right).
\end{align}
In order to avoid the alternation of signs in $\dot \Phi$, we find recursive relations with step two
\begin{align}
T_{2n + 2}  &\simeq T_{2n}  - 2c\dot \Phi _{2n},\\
\dot \Phi _{2n + 2}  &\simeq \dot \Phi _{2n}  + 2c\frac{{\dot \Phi _{2n} ^2 }}{{T_{2n} }},
\end{align}
where $c = \frac{1}{{\lambda ^ -  }} + \frac{1}{{\lambda ^ +  }}$. We are interested in finding the asymptotic behavior of $T_{2n}$ and $\dot \Phi _{2n}$. We try a solution of the form
\begin{align}
T_{2n}  &\simeq an^s,\\
\dot \Phi _{2n}  &\simeq bn^t.
\end{align}
The recursive relations at leading order become
\begin{align}
asn^{s - 1}  &=  - 2cbn^t,\\
btn^{t - 1}  &= \frac{{2cb^2 n^{2t - s} }}{a},
\end{align}
which imply that
\begin{align}
s - t &= 1,\\
 - as &= 2cb,\\
at &= 2cb.
\end{align}
The solution is
\begin{align}
s &=  - t = \frac{1}{2},\\
a &=  - 4cb.
\end{align}
$b$ in undetermined and depends on the initial values for the series, or else in the parameters of the potential. Thus the asymptotic behavior we are looking for is
\begin{align}
T_{2n}  &=  - 4cb\sqrt n,\\
\dot \Phi _{2n}  &= \frac{b}{{\sqrt n }}.
\end{align}
Using the above equations we can find asymptotic expressions for any element of the solution. Combining them we get
\begin{equation}
\dot \Phi _{2n}  =  - \frac{{4cb^2 }}{{T_{2n} }}.
\label{eq:trigasymptot}
\end{equation}

\subsection{The Stress-Energy Tensor in the Interior of the Bubble}
Now we know the asymptotic form of the solution, thus we can calculate the asymptotic form of the stress-energy tensor in the interior of the bubble. Equation (\ref{eq:trigasymptot}) implies
\begin{equation}
\left\langle {\frac{1}{2}\dot \phi ^2 } \right\rangle \sim \frac{A}{{\tau ^2 }}.
\end{equation}
Now the potential is not quadratic, which means that kinetic and potential energy do not average at the same value. Virial theorem implies
\begin{equation}
\left\langle {V\left( \phi  \right)} \right\rangle - V _ - = 2 \left\langle {\frac{1}{2}\dot \phi ^2 } \right\rangle.
\end{equation}
This means that the average value of the Lagrangian does not wash out, but asymptotically behaves as
\begin{equation}
- \left\langle \mathcal{L} \right\rangle \sim \frac{A}{{\tau ^2 }} + V _ -.
\end{equation}
If we calculate the stress-energy tensor
\begin{equation}
{T^\mu}  _\nu   = \dot \phi ^2 \frac{{x^\mu  x_\nu  }}{{ - x_\lambda  x^\lambda  }} - \mathcal{L} \delta ^\mu  _\nu,
\end{equation}
it is now going to contain except for the kinetic energy of the wall, a cosmological constant like term that depends on the size of the bubble like
\begin{equation}
\Lambda \sim \frac{A}{\tau^2} +V _ -,
\end{equation}
where $c$ depends on the specific parameters of our potential.

\subsection{A Candidate for Dark Energy}

Let's now take a wild guess. Let's suppose that big bang was the emission of a bubble, so we are actually living inside a bubble with the size of the universe. Let' also suppose that the potential of the phase transition is well fit by the triangular approximation. It actually suffices that the potential in the region of the true vacuum is triangular. The natural selection of the potential parameters is that they are of Planck scale. Finally let's suppose that the energy of the true vacuum is exactly zero.

Under these assumptions, we should today observe an effective cosmological constant of the order
\begin{equation}
\Lambda \sim \frac{M_{Pl}^2}{R_{universe}^2},
\end{equation}
which obviously depends on the age of the universe. The size of the observable universe today is about $10^{60}$ in Planck units. Thus the cosmological constant we should observe in our scenario, would be
\begin{equation}
\Lambda \sim 10^{-120} M_{Pl}^4.
\end{equation}

This is the right order of magnitude that we measure today \cite{Riess:1998cb}, \cite{Perlmutter:1998np}, \cite{Baker:1999jw}, \cite{Tegmark:2003ud}.

With this model we don't solve the original cosmological constant problem \cite{Weinberg:1988cp}, namely why the vacuum energy of SM does not gravitate. Most probably we need more information on quantum gravity to resolve this. However it resolves the cosmological constant problems that occurred after the recent measurement of it \cite{Weinberg:2000yb}. It explains its order of magnitude and if we assume that matter originates from the kinetic energy of bubble walls, then our model also explain why the matter content and dark energy content of our universe are of the same order of magnitude.

The idea of relating physical constants with cosmological quantities is not new at all. Paul Dirac in the 1930's observed that ratios of orders of magnitudes of cosmological quantities are similar to ratios of orders of magnitudes concerning the fundamental interactions. Conjecturing that this cannot be a coincidence he expressed the large number hypothesis \cite{Dirac:1937ti}, \cite{Dirac:1938mt}, \cite{Ray:2007cc}, according to which fundamental constants of nature, such as Newton's gravitational constant, depend on the age of the universe. The recent discovery of another large number, namely the ratio of the theoretical and observed vacuum energy densities, lead to similar tries to connect the energy density of the vacuum with the age of the universe \cite{nottale}, as we do in this paper.

Additionally the idea of the dark energy originating from a scalar is not new, too. Quintessence models can describe the dark energy content of the universe. In this approach the dark energy is the effect of a slow rolling scalar field instead of one that performs a dumped oscillation, as in our case. The subject of quintessence is quite broad. A nice review is given in \cite{Copeland:2007zz}.

One critical objection about our model is the singular form of the potential at the position of the true vacuum. One should be able to find a model which predicts the existence of such a vacuum and moreover develop quantum field theory in the region of such a vacuum to show that the low energy effective description is also singular. However linear potential can occur by brane interactions in an orbifold like in the ekpyrotic scenario for the big bang \cite{Moore:2000fs}, \cite{Khoury:2001wf}, \cite{Brandenberger:2001bs}. Definitely further study is required on this field.

\section{The Asymptotic Damped Oscillation Inside the Bubble}

\subsection{The Asymptotic Solution for Potential $V = a\left( {\phi  - \phi _0 } \right)^n $}

It is interesting that the Stress-Energy tensor in the interior of the bubble has a direct dependency on the size of the bubble. It is also interesting that depending on the potential this Stress-Energy tensor may represent a substance with negative pressure, thus providing a candidate for dark energy. Although it is impossible to find exact solutions for general form of the potential, we will try to calculate the asymptotic form of this damped oscillation for a potential of the form $V = a\left( {\phi  - \phi _0 } \right)^n$. In order to simplify things we assume that the potential is infinite for $\phi < \phi_0$, or else $\phi_0$ is the boundary of the configuration space. Thus the solution gets reflected when it reaches $\phi_0$. This may represent an effective field theory, where the scalar is a moduli describing geometry of some brane configuration living in an orbifold.

\begin{figure}[h]
\begin{center}
\includegraphics[angle=0,width=0.65\textwidth]{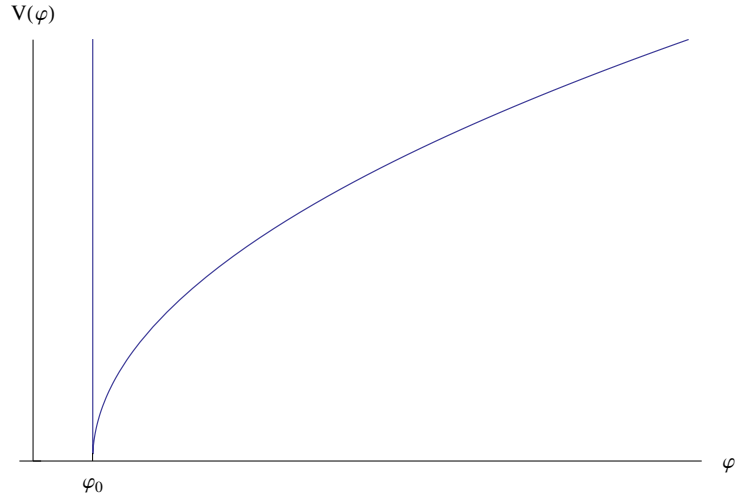}
\end{center}
\caption{The form of the potential}
\label{fig:potential}
\end{figure}

We are interested in the damped oscillations a field performs around the area of the vacuum of a potential of the form
\begin{equation}
 V\left( \phi  \right) =
 \begin{cases}
 \infty, & \phi < \phi_0 \\
 a\left( {\phi  - \phi _0 } \right)^n, & \phi > \phi_0 \\
 \end{cases}.
\end{equation}
The potential is sketched in figure \ref{fig:potential}.

Assuming a spherically symmetric solution, the equation of motion is
\begin{equation}
\ddot \phi  + \frac{3}{\tau} \dot \phi  = - V'\left( \phi  \right),
\end{equation}
where dot represents differentiation with respect to $\tau$. If we define
\begin{equation}
\frac{1}{2}\dot \phi ^2  + V\left( \phi  \right) \equiv E,
\end{equation}
the equation of motion can be written as
\begin{equation}
\dot E =  - \frac{3}{\tau }\dot \phi ^2.
\end{equation}
This form of equation gives us a better point of view about how the system dissipates. We cannot find an exact solution to the above, but we expect that when $\tau$ is large, the losses are small in one period, and thus we can express the above equation as
\begin{equation}
\left\langle {\dot E} \right\rangle  =  - \frac{6}{\tau }\left\langle T \right\rangle,
\label{eq:averageeom1}
\end{equation}
where $T \equiv \frac{1}{2}\dot \phi ^2$. Now we can use the explicit form for our potential
\begin{equation}
V = a\left( {\phi  - \phi _0 } \right)^n
\end{equation}
and the virial theorem. The latter instructs us that
\begin{equation}
\left\langle T \right\rangle  = \frac{n}{{n + 2}}\left\langle E \right\rangle ,\quad \left\langle V \right\rangle  = \frac{2}{{n + 2}}\left\langle E \right\rangle.
\end{equation}
Using the above, equation (\ref{eq:averageeom1}) can be written
\begin{equation}
\left\langle {\dot E} \right\rangle  =  - \frac{{6n}}{{n + 2}}\frac{{\left\langle E \right\rangle }}{\tau },
\end{equation}
whose solution trivially is
\begin{equation}
\left\langle E \right\rangle  = c\tau ^{ - \frac{{6n}}{{n + 2}}}.
\end{equation}
$c$ depends on the parameters of the potential. If we assume that a characteristic mass scale of the potential is $M_V$ then the above solution behaves as
\begin{equation}
\left\langle E \right\rangle  \sim \frac{{M_V ^{ - 2\frac{{n - 4}}{{n + 2}}} }}{{\tau ^{\frac{{6n}}{{n + 2}}} }}.
\end{equation}

\begin{figure}[h]
\begin{center}
\includegraphics[angle=0,width=0.65\textwidth]{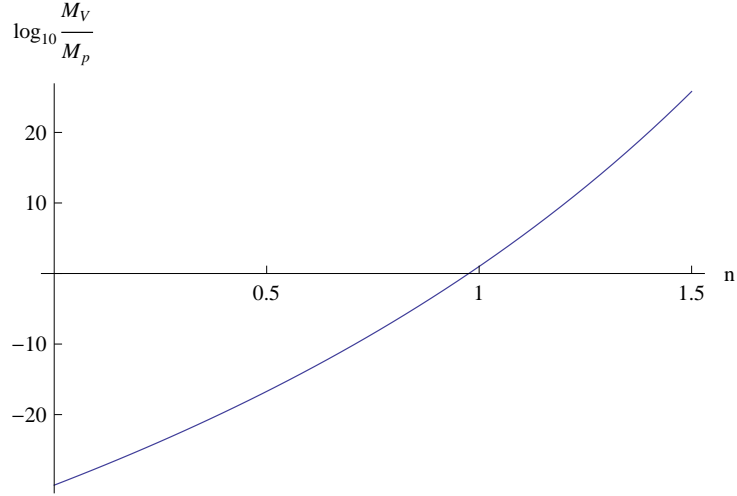}
\end{center}
\caption{The scale of the potential as function of $n$}
\label{fig:potentialscale}
\end{figure}

If we are about to explain the dark energy content of the universe as this energy stored in the damped oscillation of the scalar field, we should check what is the appropriate value for the potential mass scale $M_V$. In such case $\left\langle E \right\rangle$ has to equal the dark energy density measured today, and $\tau$ should equal the size of the observable universe $R_U$. Then $M_V$ must be
\begin{equation}
M_V  \sim \left\langle E \right\rangle ^{ - \frac{1}{2}\frac{{n + 2}}{{n - 4}}} R_U ^{\frac{{3n}}{{n - 4}}}.
\end{equation}
If we use $\left\langle E \right\rangle \equiv 10^{-120} M_p^4$ and $R_U=10^{61} l_p$ we get the behavior shown figure \ref{fig:potentialscale}.

Here we would like to notice three interesting cases.

\begin{enumerate}
\item $n=1$.
This case has already been studied in section \ref{sec:triangular}. The scale of the potential turns out to be the Planck scale, which obviously is a natural choice. Our result obviously agree with the results of the previous section.

\item $n=\frac{1}{2}$.
In this case it turns out that $M_V = 10^{-17} M_p$, thus it is about $100GeV$, the electroweak scale. It could provide a connection to other known physics.

\item $n\to 0$.
In this case the size of the universe does not help us at all to solve the hierarchy problem between the potential scale and the cosmological scale. We just convert the cosmological constant problem to a classical fine tuning problem as we need a potential with characteristic scale the cosmological one.
\end{enumerate}

\subsection{The Stress-Energy Tensor}

The stress-energy tensor is given by
\begin{equation}
{T^\mu}  _\nu   = \partial ^\mu  \phi \partial _\nu  \phi  - \mathcal{L} \delta ^\mu  _\nu.
\end{equation}

Our solution depends only on proper time, thus
\begin{equation}
\partial _\nu  \phi  = \dot \phi \left( \tau  \right)\frac{{x_\nu  }}{{\sqrt { x_\mu  x^\mu  } }}.
\end{equation}
Using the above we get
\begin{equation}
{T^\mu}  _\nu   = \dot \phi ^2 \frac{{x^\mu  x_\nu  }}{{ - x_\lambda  x^\lambda  }} -\mathcal{L} \delta ^\mu  _\nu.
\end{equation}
\begin{equation}
\left\langle {{T^\mu}  _\nu  } \right\rangle  = 2\frac{{x^\mu  x_\nu  }}{{\tau ^2 }}\left\langle T \right\rangle  - \delta ^\mu  _\nu  \left( {\left\langle T \right\rangle  - \left\langle V \right\rangle } \right).
\end{equation}

\begin{figure}[h]
\begin{center}
\includegraphics[angle=0,width=0.65\textwidth]{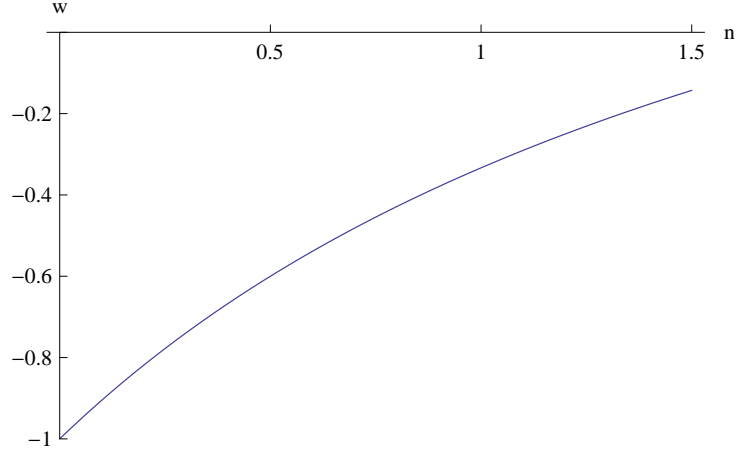}
\end{center}
\caption{$w$ as function of $n$}
\label{fig:wfactor}
\end{figure}

In the far future or in local coordinates, this is diagonal and describes an average energy density and pressure
\begin{align}
\rho  &= \left\langle T \right\rangle  + \left\langle V \right\rangle,\\
p  &= \left\langle T \right\rangle  - \left\langle V \right\rangle.
\end{align}
Use of virial theorem is adequate to calculate
\begin{equation}
w \equiv \frac{p}{\rho } = \frac{{n - 2}}{{n + 2}},
\end{equation}
which is plotted in figure \ref{fig:wfactor}.

In the previous section we distinguished three interesting cases. It turns out that the calculated $w$ for those is also interesting.

\begin{enumerate}
\item $n=1$.
In the case of a linear potential it turns out that $w = - \frac{1}{3}$. This is exactly the boundary case between an accelerating universe and a decelerating one.

\item $n=\frac{1}{2}$.
In this case we find $w = - \frac{3}{5}$. This is not the most favored value experimentally, however it corresponds to an accelerating universe.

\item$n\to 0$.
This gives us $w=-1$ which is the most favorable value for the dark energy as measured today, and indistinguishable from a cosmological constant term, as long as the equation of state is considered.
\end{enumerate}

\subsection{Inhomogeneities in the Dark Content}

We should not forget that actually the field performs an oscillation around the vacuum. That means that a fair question is whether the period of this oscillation is large enough in order to allow for measurable phenomena. If we neglect the dissipation term, we can estimate the period of the oscillation, just using conservation of $E$.
\begin{equation}
T = \sqrt 2 \int_{\phi _0 }^{\phi _0  + \left( {\frac{E}{a}} \right)^{\frac{1}{n}} } {\frac{1}{{\sqrt {E - a\left( {\phi  - \phi _0 } \right)^n } }}d\phi } =  \frac{{\sqrt {2\pi } \Gamma \left( {\frac{{n + 1}}{n}} \right)}}{{\Gamma \left( {\frac{{n + 2}}{{2n}}} \right)}}\frac{1}{{\sqrt E }}\left( {\frac{E}{a}} \right)^{\frac{1}{n}}.
\end{equation}

As the period depends on $E$ like $E^{\frac{1}{n}-\frac{1}{2}}$, for any $n$ smaller than 2, it turns out that as universe grows, the period of the oscillations is getting smaller and smaller. If we use the experimental values for dark energy density and the radius of the universe, we result in a period of oscillations at our age, that is even smaller than the Planck time, thus making these oscillations experimentally unmeasurable.

Having such high frequency oscillations seems peculiar, however we study a strange field theory, where the potential around the vacuum is not harmonic. Further study on such field theories is required.
\newpage
\section{Discussion}

We analyzed tunneling solutions for some potentials that can provide us with analytic solutions. We learned several qualitative facts about phase transitions in field theory, as well as we acquired some useful tools.

The volcano and ever better the smooth quadratic potential, can provide tools for calculating decay rates in several problems where the thin wall approximation \cite{Coleman:1977py} does not apply. Of course we already have the tools of rectangular and triangular approximation \cite{Duncan:1992ai}, however, as the decay rate per unit volume depends on the Euclidean action exponentially, a calculation of greater accuracy may be useful.

As the volcanic potential model teaches us, it is typical for such bubble solutions, that the field never reaches the true vacuum in the interior of the bubble. Instead it is performing a damped oscillation around it, whose amplitude is some kind of function of time that depends on the form of the potential around the true vacuum. This phenomenon may be of interest depending on the form of the potential.

The description of the barrier between the true and false vacuum by a non-convex potential generates questions about the uniqueness of the tunneling solution. Naively such a potential generates oscillatory behavior of the field at the region of the top of the barrier, that could result in multiple solutions. Although we don't have a proof, the smooth quadratic potential example shows that this is not the case.

An interest lesson we also learn from the smooth quadratic potential example is that the size of the emitted bubble strongly depends and it is of the same order of magnitude of the inverse of the curvature of the potential at the top of the barrier or at least the latter serves as a lower bound for the radius of the bubble. This has an interesting implication in the case of an asymptotically expanding universe. Typically phase transitions in field theory occur at temperatures of the order of characteristic quantities of the potential describing the system. Masses are such quantities. Thus according to our previous arguments the size of emitted bubbles is going to be of the order of inverse temperature. Depending on the cosmological model, the radius of the universe is also a function of temperature, thus resulting in an upper bound for the possible number of bubbles emitted. This restricts the number of bubble collisions resulting in several cosmological predictions. Of course this is not going to be the case in a post-inflationary universe.

The most interesting result is that, in a triangular potential, we observe an effective cosmological constant in the interior of the bubble, that decreases as the bubble expands, in such a way that its scale asymptotically equals the geometric means of the size of the bubble and the scale of the potential. Assuming that the parameters of the potential are of Planck scale, and that the true vacuum energy vanishes, we make a very good prediction for the cosmological constant order of magnitude, which agrees with what we measure today. Of course this effective cosmological constant term is not the only contribution to the stress-energy tensor, resulting in the ratio of pressure to energy density in the interior of the bubble being larger than the experimentally favored value -1. However, the model we use contains only one scalar and is very simplistic. The idea may be useful in the construction of a more realistic model.

Later we extend our analysis to study the damped oscillation for a more general potential. It turns out that other interesting options also exist. If the potential is proportional to the square root of the field, then a potential with characteristic parameters of the electroweak scale predicts a dark energy content of the right order of magnitude and an accelerating universe.

Of course such a model for the cosmological constant predicts a vacuum energy that depends on the distance from the bubble wall, thus on position inside the bubble. Current experiments do not rule out such a dependence, thus this is an direct experimental prediction of our model.

Such a time dependent cosmological constant is also predicted by quintessence models. The advantage of our approach is that the value of the observed cosmological constant is related with the size of the universe in a more direct way.

Moreover even in the case of a quadratic potential there is still some non-trivial form for the stress-energy tensor that could have interesting cosmological implications in the expansion of the universe. In the case of more singular potentials, a connection with inflation could also be interesting. Finally there is the open direction of generalizing to different potentials, or greater number of fields. Interesting behavior like the one we discovered in the triangular model, may appear in different cases. We believe that the most interesting future direction would be the inclusion of gravity into the problem. This way, we would be able to directly observe the effects of the discovered phenomena in the evolution of the universe.

\section*{Acknowledgments}

I would like to especially thank D. Blas, L. Motl, K. Papadodimas and R. Rattazzi for useful discussions.

\end{document}